\documentclass[aps,preprint]{revtex4}
\usepackage{amsfonts}
\usepackage{amsmath}
\usepackage{amssymb}
\usepackage{mathrsfs}
\usepackage{subfigure}
\usepackage{graphicx}
\usepackage{epstopdf}
\usepackage{float}
\usepackage{color}

\begin{document}
\preprint{CTP-SCU/2021030}
\title{Shadow and Photon Sphere of Black Hole in Clouds of Strings and Quintessence}
\author{Aoyun He}
\email{heaoyun@stu.scu.edu.cn}
\author{Jun Tao}
\email{taojun@scu.edu.cn}
\author{Yadong Xue}
\email{xueyadong@stu.scu.edu.cn}
\author{Lingkai Zhang}
\email{zhanglingkai@stu.scu.edu.cn}
\affiliation{Center for Theoretical Physics, College of Physics, Sichuan University, Chengdu, 610065, China}

\begin{abstract}
In this work, we study the shadow and photon sphere of the black bole in clouds of strings and quintessence with static and infalling spherical accretions. We obtain the geodesics of the photons near a black hole with different impact parameters $b$ to investigate how the string clouds model and quintessence influence the specific intensity by affecting the geodesic and the average radial position of photons. And the range of string clouds parameter $a$ is constrained to ensure that the shadow can be observed. Moreover, the light sources in the accretion follow a normal distribution with an attenuation factor $\gamma$ and we use a model of the photon emissivity $j(\nu_e)$ to get the specific intensities.  Furthermore, the shadow with static spherical accretion is plotted, which shows the apparent shape of the shadow is a perfect circle and the value of $\gamma$ affects the brightness of the photon sphere. Then, we investigate the profile and specific intensity of the shadows with static and infalling spherical accretions respectively. The interior of the shadows with an infalling spherical accretion will be darker than that with the static spherical accretion, and the specific intensity with both static and infalling spherical accretion gradually converges.
\end{abstract}

\maketitle

\section{Introduction}

The Black Hole's apparent shape from a distant observer when illuminated by a background source of light is referred to as a black hole shadow. When a shadow appears, the light, or photons, can orbit around this spherically symmetric black hole at a constant radius which form a bright ring called the Photon Sphere. Although such an orbit is unstable, it is nevertheless important from a physical viewpoint because it defines the boundary for a BH shadow between capture and non-capture of a cross-section of light rays by the black hole. This boundary has played an important role in determining, for example, the optical appearance of a black hole with thin accretions\cite{Luminet:1979nyg}.  

The Event Horizon Telescope (EHT) collaboration which is now the world's biggest astronomy project reported 1.3 mm Very Long Baseline Interferometry (VLBI) observations of the nucleus of the nearby galaxy M87, achieving angular resolution comparable to the expected size of the supermassive black hole \cite{Akiyama:2019bqs, Akiyama:2019brx, Akiyama:2019cqa, Akiyama:2019eap, Akiyama:2019fyp, Akiyama:2019sww}. It promotes the current theoretical study of black hole based on modern astronomical observation methods. This project also corroborates the directly observations to the black hole shadows. Further, the shape of a shadow could be used to study gravity near the event horizon and find whether the general relativity is consistent with the observations \cite{Mizuno:2018lxz} which enables a direct probe of the General Relativity (GR) in the extreme environment.

The shadow of the Schwarzschild black hole was first discussed in \cite{Synge:1966okc}. Bardeen \cite{Bardeen:1972fi} soon studied the shadow cast by the Kerr black hole. In recent years, this topic has been extended to many other black holes by various researchers \cite{Hioki:2009na, Takahashi:2005hy, Gan:2021xdl, Tsukamoto:2017fxq, Wei:2013kza, Abdujabbarov:2012bn, Schee:2008kz, Podolsky:2012he, Bambi:2011yz, Bambi:2010hf, Younsi:2016azx, Cunha:2016wzk, Belhaj:2021rae, Saha:2018zas, Eiroa:2017uuq, Amir:2016cen, Abdujabbarov:2016hnw, Lara:2021zth}. Multiple shadows of a single black hole have also been discussed \cite{Cunha:2015yba, Grover:2018tbq}, as well as the shadow of multiple black holes \cite{Yumoto:2012kz}. The black hole shadow in modified GR has also been investigated in \cite{Amarilla:2010zq, Kumar:2017tdw, Mureika:2016efo}. The models of black hole considering the accretion which can modify the silhouette are studied in \cite{Perlick:2017fio, Cunha:2018acu, Konoplya:2019sns}. 

We consider the black hole with dark energy as the Standard Model of cosmology suggests that dark energy is dominant in our universe \cite{Pope:2004cc, Komatsu:2010fb}and its dynamic may affect the black hole \cite{Kiselev:2002dx} whose effect on spacetime is similar to the cosmological constant or vacuum energy \cite{Copeland:2006wr}. Moreover, modern astronomical observers have found that the universe is in accelerating expansion \cite{SupernovaCosmologyProject:1998vns, SupernovaSearchTeam:1998fmf, SupernovaSearchTeam:1998cav}, implying a state of negative pressure. And the quintessence dark energy is one of the candidates to interpret the negative pressure \cite{Vagnozzi:2020quf}. In this model, the state equation of the pressure is $p=\omega_q\rho _q$, where $\rho _q$ is the energy density and $\omega_q$ is the state parameter which satisfies $-1<\omega_q<-1/3$ \cite{Stern:1999ta, Bahcall:1999xn, Steinhardt:1999nw, Wang:1999fa}. The effect of an accelerated expanding universe on the shadow of a black hole is studied in \cite{Frion:2021jse}. It is also natural to study the effect of the quintessence dark energy \cite{Kiselev:2002dx}, and the investigation can bring new insights and impose restrictions on the quintessence model \cite{Nam:2019rop}. Shadows affected by other dark energy models are studied in \cite{Zeng:2020vsj, Khan:2020ngg}.

Theoretical developments propose that the basic unit of nature is one-dimensional strings instead of point-like particles. Studying Einstein's equations with string clouds may be critical because relativistic strings can be used to construct suitable models \cite{Letelier:1979ej}. A cloud of strings as the source of the gravitational field was first considered by Latelier, who found an exact solution of Schwarzschild BH surrounded by a cloud of strings \cite{Letelier:1979ej, Letelier:1983vka}. Later, the rotating BH with a cloud of strings was investigated \cite{Barbosa:2016lse, Toledo:2020xnt}. Then, the study of a cloud of strings was extended to modified theories of gravity such as Lovelock gravity \cite{Herscovich:2010vr, Toledo:2019szg}. Recently, the exact solution of Schwarzschild black hole surrounded by a cloud of strings in Rastall gravity is obtained in \cite{Cai:2019nlo}.

In this paper, we study the shadow and photon spheres of the black hole surrounded by a cloud of strings and quintessence with both static and infalling spherical accretions. The black hole thermodynamics combined effects of the string clouds and quintessence are considered in \cite{Chabab:2020ejk}. Afterwards, studies in this area were proposed for charged AdS black hole \cite{Toledo:2019amt} as well as Lovelock gravity \cite{deMToledo:2018tjq}. Furthermore, we explore how the photons emissivity $j(\nu _e)$ affects the shadow of the black hole.

This paper is organized as follows. In Section II, we give the metric of a black hole surrounded by quintessence and clouds of strings. Then, we derive the complete null geodesic equations for a photon moving around the static black hole. In addition, the range of parameters affected by string clouds is discussed. In Section III, we study the shadows and photon spheres with static spherical accretions of the static black hole. In Section IV, the black hole shadows and photon spheres with an infalling spherical accretion are investigated. In Section V,  we discuss and conclude our results.

\section{Metric and photon Geodesic}

The metric of a uncharged static black hole surrounded by quintessence and clouds of strings is given by \cite{Chabab:2020ejk}
\begin{align}
    ds^2=-f(r)dt^2+\frac{1}{f(r)}dr^2+r^2(d\theta ^2+sin^2\theta d\phi ^2),
    \label{Metric of spherical black hole}
\end{align}
with
\begin{align}
    f(r)=1-a-\frac{2M}{r}-\frac{\alpha }{r^{3\omega_q+1}},
    \label{General form of f(r)}
\end{align}
where $a$ is the constant affected by the cloud of strings, $M$ is the mass of the black hole, $\omega_q$ is the state parameter of quintessence and $\alpha$ is a constant related to the quintessence with density $\rho_q$ as
\begin{align}
    \rho_q=-\frac{\alpha}{2}\frac{3\omega_q }{r^{3(\omega_q+1)}}.
    \label{Density of quintessence}
\end{align}

The pressure of the quintessence dark energy $p$ should be negative due to the cosmic acceleration. The range of $\omega_q$ is $-1 < \omega_q < -1/3$ for $p=\omega_q \rho_q$ \cite{Stern:1999ta, Bahcall:1999xn, Steinhardt:1999nw, Wang:1999fa}, which causes the existence of event horizon and cosmological horizon. And the region between the two event horizons is called the outer communication domain, because any two observers in it can communicate with each other without being blocked \cite{outer communication, outer communication topology}.

Setting $f(r)=0$ to get two roots, the smaller one is the radius of event horizon $r_h$ and the another is the cosmological horizon $r_c$. In general, it is not easy to get analytic solutions, but for $\omega_q=-2/3$, the analytical solutions of $f(r)=0$ can be written as
\begin{align}
    r_h & =\frac{1 - a + \sqrt{(1-a)^2 - 8 M \alpha}}{2\alpha}, \\
    r_c & =\frac{a - 1 + \sqrt{(1-a)^2 - 8 M \alpha}}{2\alpha}.
    \label{Horizon of -2/3}
\end{align}
The numerical values of $r_c$, $r_h$ corresponding to other $\omega_q$ are listed in Table. \ref{Table}.

Then, we calculate the geodesic of photons to investigate the light deflection with the Euler-Lagrange equation. The Lagrangian takes the form as
\begin{align}
    \mathcal{L} =\frac{1}{2}g_{\mu v}\dot{x}^{\mu}\dot{x}^{v},
    \label{Lagrangian of photon}
\end{align}
where the dot represents the derivative to the affine parameter.

Due to the spherical symmetry of metric, it is convenient to study the photon trajectories on the equatorial plane with the initial condition $\theta=\pi/2$ and $\dot{\theta}=0$. Combining Eqs. (\ref{Metric of spherical black hole}) and (\ref{Lagrangian of photon}) with the Euler-Lagrange equation, we get the expressions of time, azimuth, and radial components of the four velocities. Since the metric does not explicitly depend on time $t$ and azimuthal angle $\phi$, there are two corresponding conserved quantities $E$ and $L$. Then, by redefining the affine parameter $\lambda$ as $\lambda/{\left\lvert L\right\rvert }$ and using the impact parameter $b={\left\lvert L\right\rvert}/E$ \cite{Zeng:2020dco}, we get the following equations
\begin{align}
    \dot{t}       & =\frac{1}{bf(r)},
    \label{Derivative of t}                                                \\
    \dot{\phi}    & =\pm \frac{1}{r^2},
    \label{Derivative of phi}                                              \\
    \frac{1}{b^2} & =\dot{r}^2+\frac{1}{r^2}f(r)  \label{Derivative of r},
\end{align}
where $+$ and $-$ in Eq. (\ref{Derivative of phi}) represent the counterclockwise and clockwise direction, respectively for the motion of photons. Moreover, Eq. (\ref{Derivative of r}) can be rewritten as
\begin{align}
    \dot{r}^2+V(r)=\frac{1}{b^2},
    \label{Relationship between velocity and effective potential}
\end{align}
where $V(r)=1/{r^2}f(r)$ is the effective potential. At the photon sphere, the photon trajectories satisfy $\dot{r}=0$ and $\ddot{r}=0$, which means
\begin{align}
    V(r)=\frac{1}{b^2},\ \ V'(r)=0.
    \label{Feature of photon sphere}
\end{align}
In general, for the radius of photon sphere $r_{ph}$ and corresponding impact parameter $b_{ph}$, it is not easy to obtain the analytic results. But for $\omega_q=-2/3$, they can be expressed as
\begin{align}
    r_{ph} & =\frac{1-a-\beta}{\alpha},                                                        \\
    b_{ph} & =\sqrt{-\frac{\left(\beta +a-1\right)^3}{\alpha ^2 \left(-a \left(\beta +a-2\right)+\beta +4 \alpha
            M-1\right)}},
\end{align}
where
\begin{align}
    \beta=\sqrt{(1-a)^2-6 \alpha M}.
    \label{beta}
\end{align}
In order to obtain a realistic solution, there are restrictions $(1-a)^2\geqslant 6\alpha M$ between $a$ and $\alpha$ with $\omega_q=-2/3$. Similar numerical results can be obtained for other $\omega_q$.

The data of $r_{ph}$, $b_{ph}$, $r_h$ and $r_c$ is listed in Table. \ref{Table} with $\alpha = 0.01$, $M=1$ and $a=0.1$. The $b_{ph}$, $r_h$ and $r_c$ are monotonic while $r_{ph}$ is non-monotonic with the decrease of $\omega_q$. The event horizon radius $r_c$ approaches the cosmological horizon radius $r_h$ with the decrease of $\omega_q$, which leads to a narrower domain of the outer communication. It is worth noting that when $\omega_q=-1/3$, the cosmological horizon radius $r_c$ does not exist, which we use - to denote.
\begin{table}[htbp]
    \begin{center}
        \begin{tabular}[b]{cccccccccc}
            \hline
            $\omega_q$ & -1/3    & -0.4             & -0.5    & -0.6    & -2/3    & -0.7    & -0.8    & -0.9    & -1      \\
            \hline
            $r_{ph}$   & 3.37079 & 3.3764           & 3.38524 & 3.39346 & 3.39746 & 3.39860 & 3.39602 & 3.37792 & 3.33333 \\
            $b_{ph}$   & 5.37669 & 6.21753          & 6.27757 & 6.36711 & 6.45119 & 6.50322 & 6.71598 & 7.06308 & 7.66965 \\
            $r_h$      & 2.04402 & 2.25165          & 2.25997 & 2.27085 & 2.27998 & 2.28525 & 2.30464 & 2.33145 & 2.37016 \\
            $r_c$      & -       & $5.9\times 10^9$ & 8095.55 & 274.403 & 87.7200 & 57.6865 & 23.1518 & 12.5872 & 8.07703 \\
            \hline
        \end{tabular}
        \caption{The data of $r_{ph}$, $b_{ph}$, $r_h$ and $r_c$ for different $\omega_q$ with $M = 1$, $\alpha = 0.01$ and $a=0.1$.}
        \label{Table}
    \end{center}
\end{table}

\begin{figure}[htbp]
    \begin{center}
        \subfigure[V(r)]{
            \includegraphics[width=5.5cm]{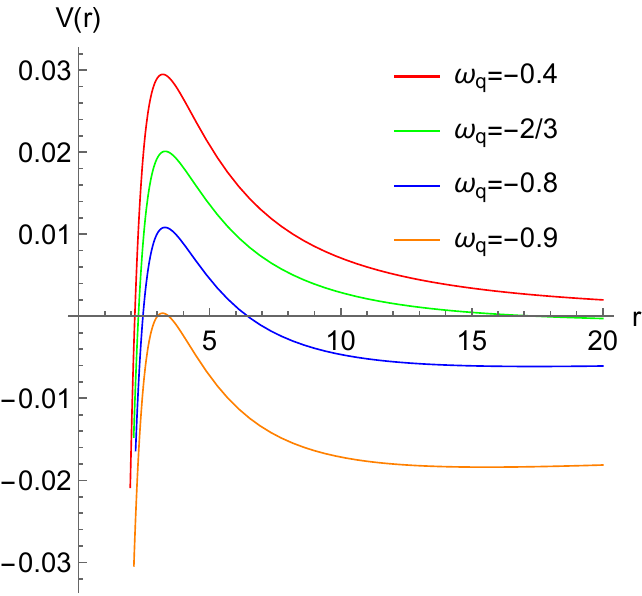}
        \label{V(r)}}
        \qquad
        \subfigure[$\omega_q =-0.4$]{
            \includegraphics[width=5.5cm]{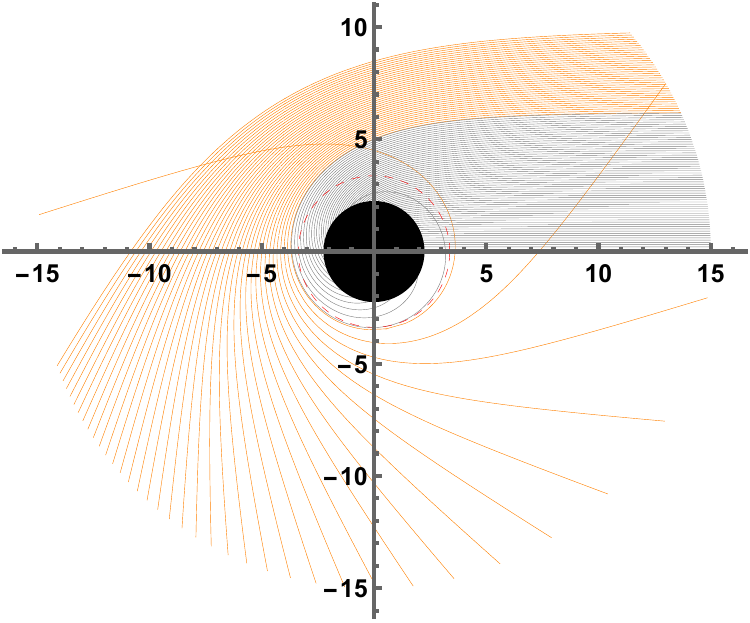}
        \label{Geo-04}}\\
        \subfigure[$\omega_q =-2/3$]{
            \includegraphics[width=6cm]{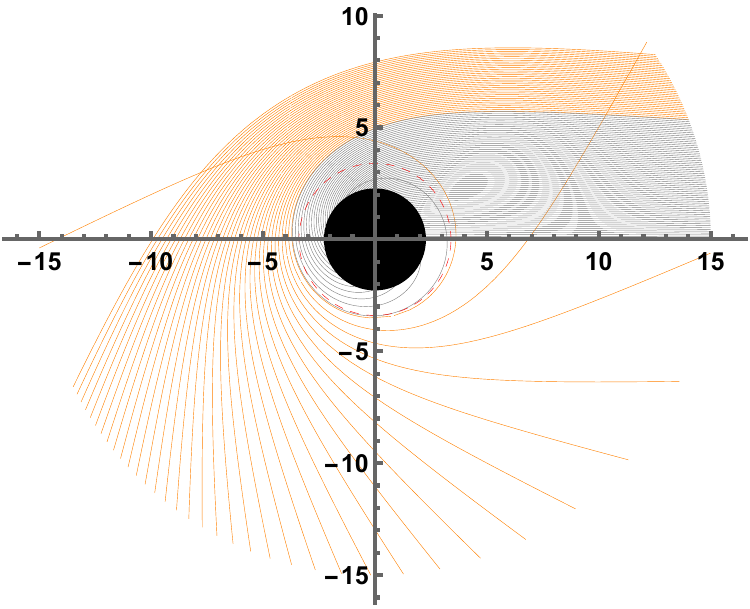}
        \label{Geo-0667}}
        \qquad
        \subfigure[$\omega_q =-0.9$]{
            \includegraphics[width=5.5cm]{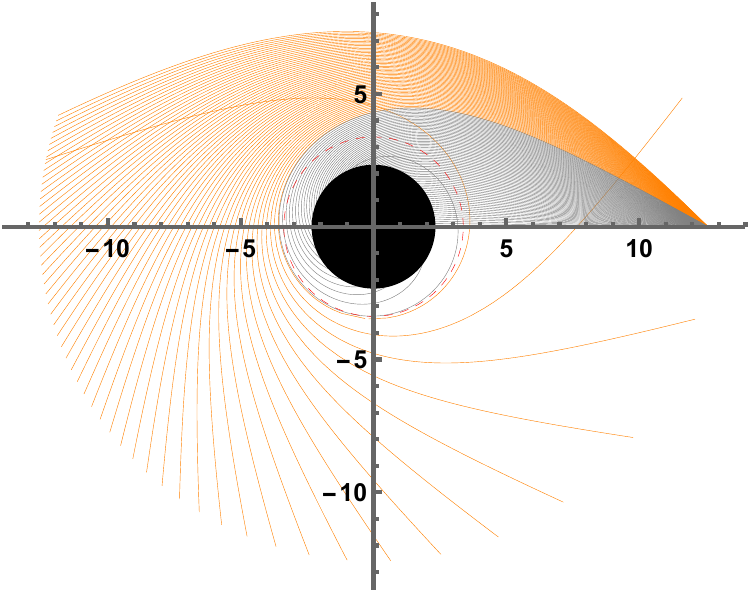}
            \label{Geo-09}
        }
        \caption{The effective potential $V(r)$ in Fig. \ref{V(r)} and the trajectories of the photons in other images for different $\omega_q$ with $\alpha =0.01$ and $a=0.1$.}
        \label{Trajectories of light rays}
    \end{center}
\end{figure}

The Eq. (\ref{Relationship between velocity and effective potential}) shows that the null geodesic depends on the impact parameter $b$ and the effective potential $V(r)$. The effective potential for different $\omega_q$ is plotted in Fig. \ref{V(r)}. The effective potential vanishes at the event horizon, increases and reaches a maximum at the photon sphere, finally vanishes at the cosmological horizon. 

The geodesics of photons can be plotted by the equation of motion. Combining Eqs. (\ref{Derivative of phi}) and (\ref{Derivative of r}), we have
\begin{align}
    \frac{dr}{d\phi}=\pm r^2\sqrt{\frac{1}{b^2}-\frac{1}{r^2}f(r)}\equiv \Phi(r).
    \label{Light ray}
\end{align}

For a photon that is approaching the black hole, it either escapes or falls into the black hole. When $b > b_{ph}$, it escapes and we denote the nearest radial position of the trajectory as $r_i$. The turning point $r_i$ can be obtained from the equation $\Phi(r) = 0$. The radial position $r_i$ is important to derive the photons trajectory, and it can also be the lower integral limit in the backward ray shooting method \cite{Luminet:1979nyg}. For $b < b_{ph}$, the photons trajectories would not have $r_i$ because the photon will continue to approach the black hole and finally falls into it. To get the trajectories of the photons, the location of the observer is also important. Physically, the observer is located near the cosmological horizon in the outer domain of communication. In other words, it will start from a position near $r_c$.

Setting the starting point of the photons trajectories on the x-axis close to $r_c$ and our plot range to be $r\leqslant 15$, the trajectories of photons are plotted in Fig. \ref{Trajectories of light rays} for different quintessence state parameter $\omega_q$ with $a=0.1$ and $\alpha=0.01$ according to Eq. (\ref{Light ray}) and the analysis above. 

The quintessence state parameter $\omega_q$ will impact the location of the starting point and the curvature of geodesics. The values of $r_c$ for different $\omega_q$ are listed in Table. \ref{Table}. For $\omega_q=-0.4$ in Fig. \ref{Geo-04}, the geodesics are near-parallel because the value of $r_c$ is much larger than our plot range ($r_c=5.9\times 10^9$). Since the cosmological horizon $r_c=12.5872$ from Table. \ref{Table} when $\omega _q=-0.9$, the plot range should be $r\leq r_c$ in Fig. \ref{Geo-09}. The geodesics here are more curved since the value of $r_c$ is close to the radius of cosmological horizon $r_h$ ($r_h=2.33145$).  

For the geodesics in Fig. \ref{Trajectories of light rays}, the orange and gray lines represent the geodesics of photons starting from $b>b_{ph}$ and $b<b_{ph}$ respectively. It is worth noting that when $b=b_{ph}$, the photons will continue to rotate around the black hole in an unstable circular orbit \cite{Teo:2020sey} locating at $r_{ph}$, which is also known as the photon sphere \cite{Black Hole Shadows} and the red dashed line represents the photon sphere and the black disk represents the black hole. 

Next, we study the geodesics for different $a$ with fixed $\omega_q$ and $\alpha$. The values of $r_{ph}$, $b_{ph}$, $r_h$ and $r_c$ are also listed in Table. \ref{Table 2} for the following analysis. Using the same method, we set $\omega_q =-2/3$ and $\alpha=0.01$ to study geodesics under the influence of the string clouds by calculating the values of various radius and get the trajectories of photons under different $a$.

\begin{table}[H]
    \begin{center}
        \begin{tabular}[b]{cccccccc}
            \hline
            $a $     & 0       & 0.1     & 0.2     & 0.3     & 0.4     & 0.5     & 0.6     \\
            \hline
            $r_{ph}$ & 3.04640 & 3.39746 & 3.84227 & 4.42561 & 5.22774 & 6.41101 & 8.37722 \\
            $b_{ph}$ & 5.44501 & 6.45119 & 7.82587 & 9.80258 & 12.864  & 18.2114 & 30.0948 \\
            $r_h$    & 2.04168 & 2.27998 & 2.58343 & 2.98438 & 3.54249 & 4.38447 & 5.85786 \\
            $r_c$    & 97.9583 & 87.7200 & 77.4166 & 67.0156 & 56.4575 & 45.6155 & 34.1421 \\
            \hline
        \end{tabular}
        \caption{The values of $r_{ph}$, $b_{ph}$, $ r_h$ and $ r_c$ for different $a$ with $\omega_q =-2/3$ and $\alpha = 0.01$.}
        \label{Table 2}
    \end{center}
\end{table}

\begin{figure}[H]
    \begin{center}
        \subfigure[$a=0$]{
            \includegraphics[height=5.5cm]{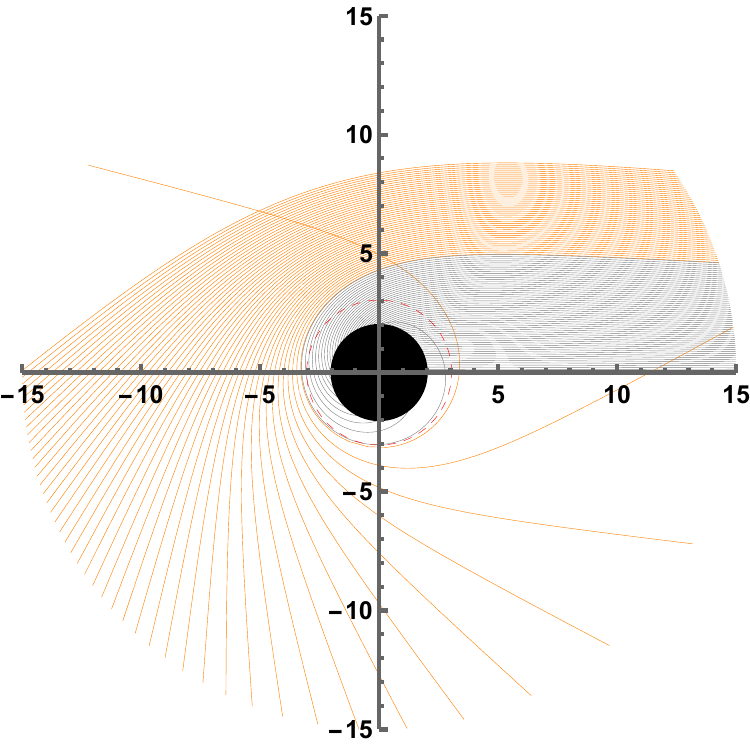}
        \label{Geo0}}
        \qquad
        \subfigure[$a=0.1$]{
            \includegraphics[height=5.5cm]{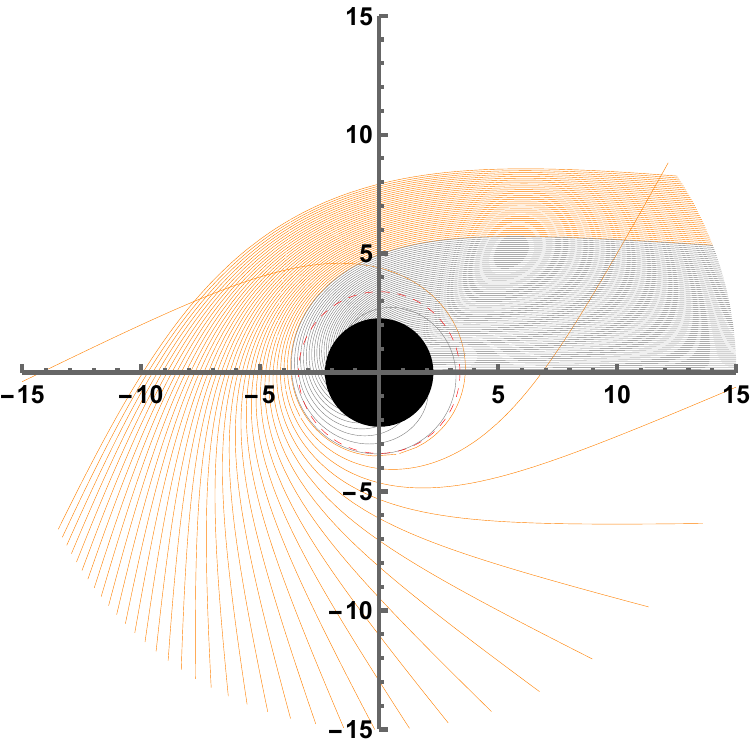}
        \label{Geo01}}\\
        \subfigure[$a=0.3$]{
            \includegraphics[height=5.5cm]{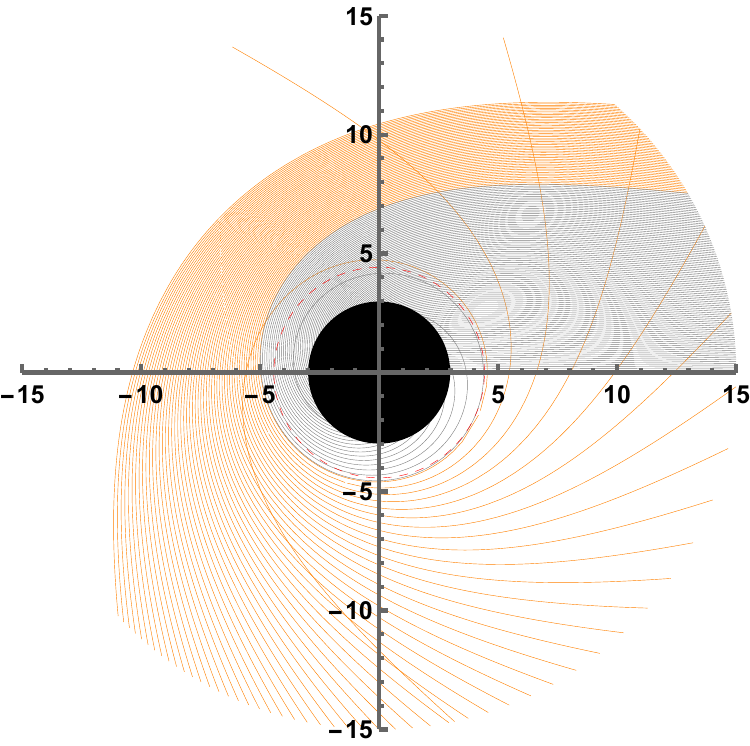}
        \label{Geo03}}
        \qquad
        \subfigure[$a=0.6$]{
            \includegraphics[height=5.5cm]{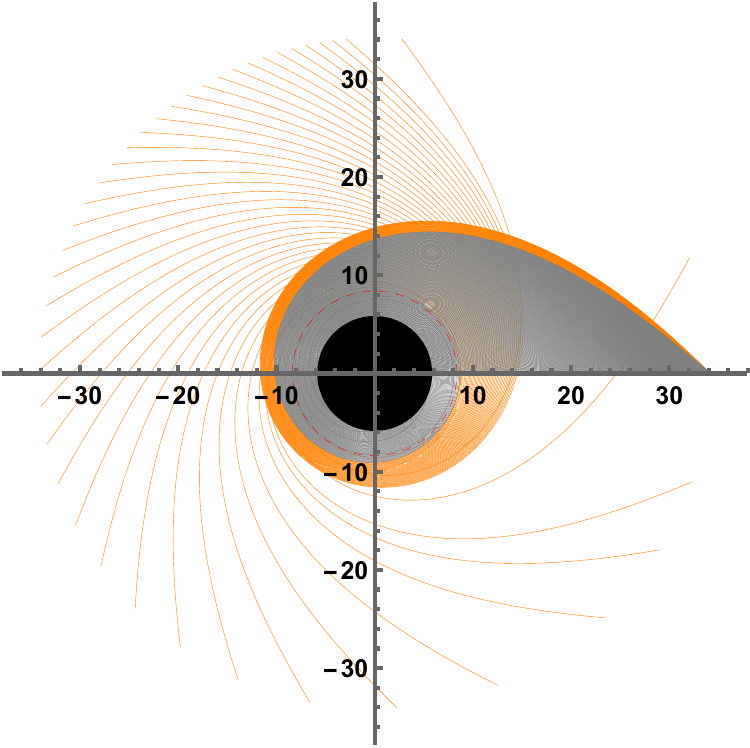}
        \label{Geo06}}
        \caption{The trajectory of the photons for different $a$ with $\omega_q=-2/3$ and $\alpha= 0.01$.}
        \label{Trajectories of light rays 2}
    \end{center}
\end{figure}

From Table. \ref{Table 2}, the $r_{ph}$, $b_{ph}$, $r_h$ and $r_c$ all vary monotonically with $a$. The geodesics with different $a$ are drawn in Fig. \ref{Trajectories of light rays 2} either. According to the value of $r_c$ in Table. \ref{Table 2}, the same conclusion can be obtained as in the above analysis. In Fig. \ref{Geo06}, the plot range is $r\leqslant r_c$ ($r_c=34.1421$). it can be seen that the deflection angle of the photons increases with increasing $a$. 

The range of $a$ is not arbitrary, it should be noted that the cosmological horizon $r_c$ must be larger than the black hole horizon $r_h$ to ensure that the outer communication domain exists. With the increase of $a$, the distance between $r_c$ and $b_{ph}$ reduces, which suggests the parameter $a$ has an upper limit $a_m=a_m(\omega_q)$ to have $r_c > b_{ph}$. For example, when $\omega _q=-2/3$ and $\alpha =0.01$, we get $a<0.612158$.

\section{Shadows and photon spheres with static spherical accretions}

In this section, we study the effect of accretion profile on black hole shadows and take spherical accretion as an example. Based on the backward ray shooting method \cite{Luminet:1979nyg} we focus on the specific intensity observed by the static observer \cite{Jaroszynski:1997bw, Bambi:2013nla}. The photons emissivity can be expressed as
\begin{align}
    j(\nu_e)\varpropto \rho(r) P(\nu_e),
    \label{jjj}
\end{align}
where $\rho(r)$ is the density of light sources in the accretion and $P$ represent the probability about spectral distribution. For the spherical accretions, the density of light sources follows a normal distribution and can be expressed as
\begin{align}
    \rho(r)=\sqrt{\frac{\gamma}{\pi}}e^{-\gamma r^2},
    \label{density}
\end{align}
where $\gamma$ is the coefficient to control the decay rate and $\sqrt{{\gamma}/{\pi}}$ is the normalization factor.

The light emitted by the source in the accretion is not strictly monochromatic, which obeys a Gaussian distribution \cite{Luminet:1979nyg}. We assume the central frequency is $\nu_*$ and its width is $\Delta\nu$, while the effect of photons outside the spectral width is neglected in this paper. Then the probability $P$ that the frequency of photon is located at the spectrum width can be written as
\begin{align}
    P(\nu_e)=\int_{\nu_*-\Delta\nu}^{\nu_*+\Delta\nu} \frac{1}{\Delta\nu\sqrt{\pi}}e^{-\frac{(\nu-\nu_*)^2}{\Delta\nu^2}} \,d\nu,
    \label{H}
\end{align}
where $1/{\Delta\nu\sqrt{\pi}}$ is the normalization factor. Combining Eqs. (\ref{jjj}), (\ref{density}) and (\ref{H}), we can express photons emissivity in the rest-frame of the emitter as
\begin{align}
    j(\nu _e)\propto \int_{\nu_*-\Delta\nu}^{\nu_*+\Delta\nu} \frac{\sqrt{\gamma}}{\pi\Delta\nu}e^{-\frac{(\nu-\nu_*)^2}{\Delta\nu^2}-\gamma r^2}\,d\nu.
    \label{jv}
\end{align}

The specific intensity received by a distant observer is given as an integral along the null geodesic \cite{Luminet:1979nyg}, and can be expressed as 
\begin{align}
    I(b)=\int_{ray}\frac{\nu_o^3}{\nu_e^3}j(\nu_e)dl_{prop} = \int_{ray}g^3j(\nu_e)dl_{prop},
    \label{Intensity}
\end{align}
where $\nu _o$ is the frequency observed and $dl_{prop}$ is the proper length as measured in the frame comoving with acceleration, the ratio of $\nu_o$ and $\nu_e$ is the redshift factor $g$. For a static spherically symmetric black hole, we have $g=\sqrt{f(r)}$. And in this spacetime, we can easily obtain
\begin{align}
    dl_{prop}=\pm \sqrt{f(r)^{-1}+r^2(\frac{d\phi}{dr})^2}dr,
    \label{lpp}
\end{align}
where $+$ and $-$ correspond to the case that the photon approaches and leaves the black hole respectively.
\begin{figure}[H]
    \begin{center}
        \includegraphics[width=10cm]{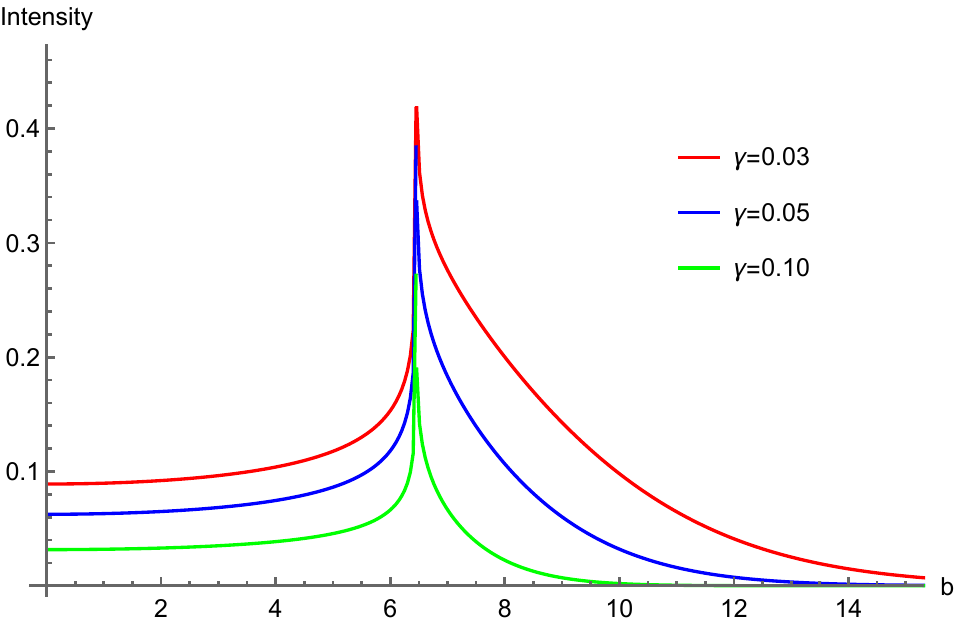}
        \caption{The specific intensity with static spherical accretion for different $\gamma$ with $\omega_q=-2/3$, $a=0.1$ and $\alpha=0.01$.}
        \label{Graphic of Gamma}
    \end{center}
\end{figure}

Combining the Eqs. (\ref{jv}), (\ref{Intensity}) and (\ref{lpp}), we can get the specific intensity $I(b)$ with impact parameter $b$. Then use it to investigate different coefficient $\gamma$ with static spherical accretion in Fig. \ref{Graphic of Gamma}. The coefficient $\gamma$ affects the intensity's decay rate. As $b$ increases, the intensity ramps up first when $b<b_{ph}$, then reaches a peak at $b_{ph}$, and finally drops rapidly to the bottom. For $b=b_{ph}$, the photon revolves around the black hole several times, causing the observed intensity to be maximal. However, due to the limitation of calculation accuracy and the logarithmic form of the intensity, the actual calculated intensity will never reach infinity \cite{Black Hole Shadows}. For $b>b_{ph}$, the density of light sources in the accretion reduces, and then the observed intensity vanishes for large enough $b$. So the coefficient $\gamma$ does not impact the intrinsic properties of spacetime geometry, which means the peak of the specific intensity is always located at $b=b_{ph}$. For convenience, the images of specific intensity below are all fixed with $\gamma=0.03$.

\begin{figure}[H]
    \begin{center}
        \subfigure[Intensity]{
            \includegraphics[width=6.5cm]{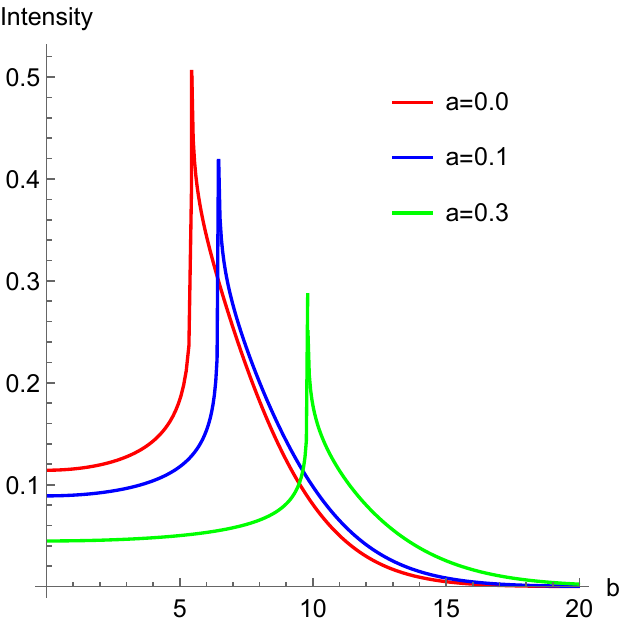}
        \label{ina}}
        \qquad
        \subfigure[a=0]{
            \includegraphics[width=6.5cm]{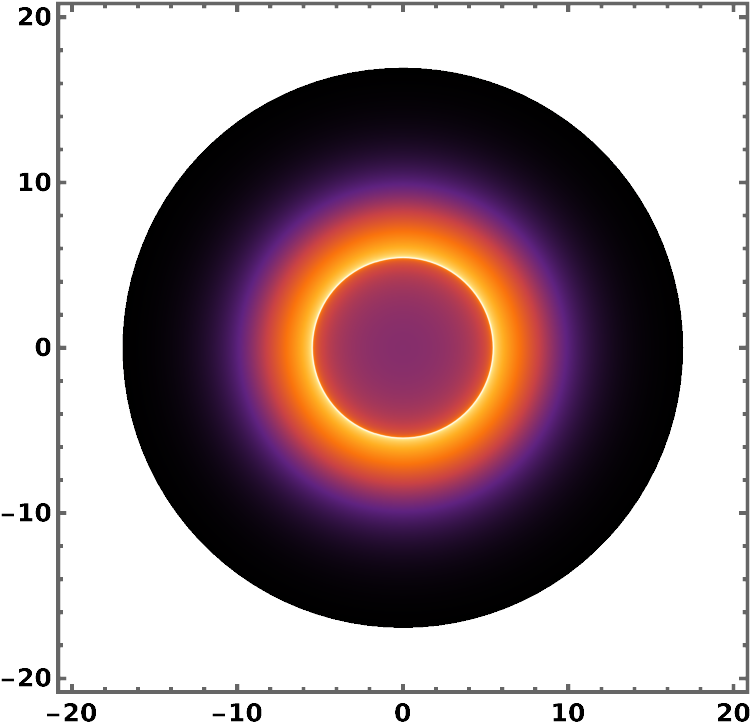}
        \label{Shadow0}}\\
        \subfigure[a=0.1]{
            \includegraphics[width=6.5cm]{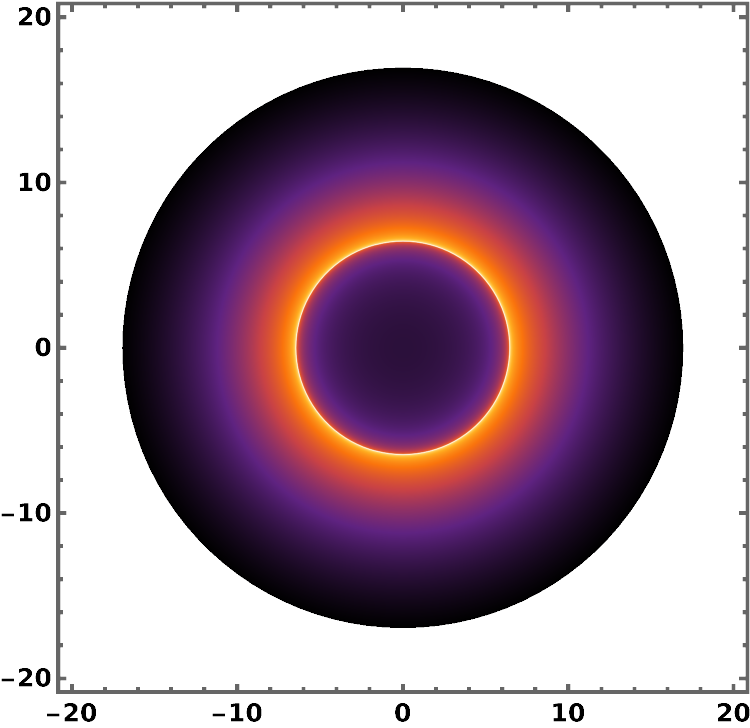}
        \label{Shadow01}}
        \qquad
        \subfigure[a=0.3]{
            \includegraphics[width=6.5cm]{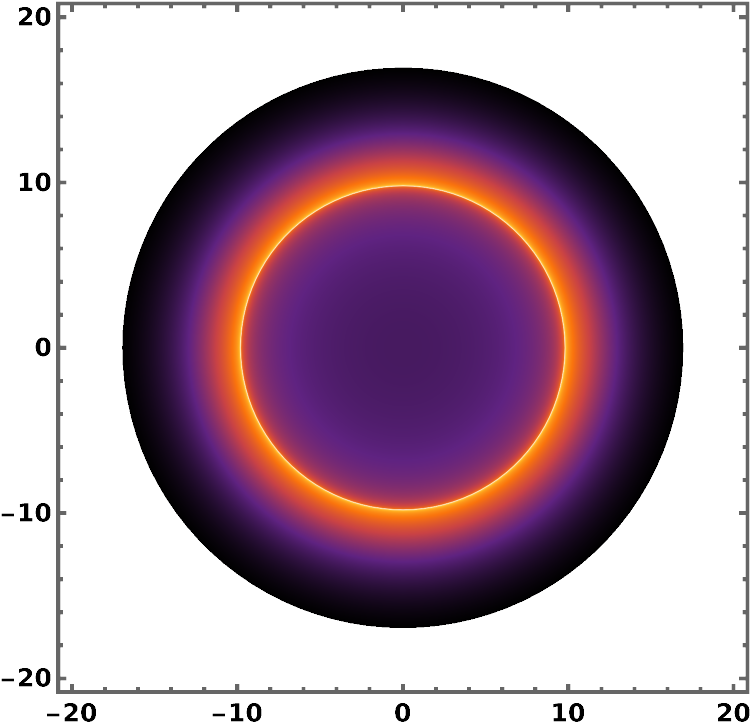}
        \label{Shadow03}}
        \caption{The specific intensity for different $a$ in Fig. \ref{ina} and the black hole shadow with static spherical accretion for different $a$ in other images with $\omega_q = -2/3$ and $\alpha = 0.01$.}
        \label{Observed specific intensities}
    \end{center}
\end{figure}

The observed specific intensities and shadows for different parameter $a$ are plotted in Fig. \ref{Observed specific intensities} with $\omega_q=-2/3$ and $\alpha =0.01$. For the specific intensity in Fig. \ref{ina}, the string clouds parameter $a$ affects the maximum value of specific intensity and the radius of the photon sphere, which is consistent with the analysis in Table. \ref{Table 2}. As the parameter $a$ increases, the radius of the cosmological horizon $r_c$ decreases rapidly which causes the upper limit of the integration to reduce and the lower limit of the integration $r_h$  to get larger, albeit slowly. The result is that the geodesics of photons shortens rapidly with $a$ increasing which leads to a smaller average specific intensity.

Fig. \ref{Observed specific intensities} show the black hole shadows for different parameter $a$ in the region $b\leqslant 17$. Different colors correspond to different values of the specific intensity, and we show all profiles of the shadows with one color function in which the greater specific intensity means the brighter, while the smaller means the darker. We can directly compare the intensity magnitude inside and outside the photon sphere. The shadow is circularly symmetric, and the photon sphere is a bright ring outside the black hole.
\begin{figure}[H]
    \begin{center}
        \subfigure{
            \includegraphics[height=5.5cm]{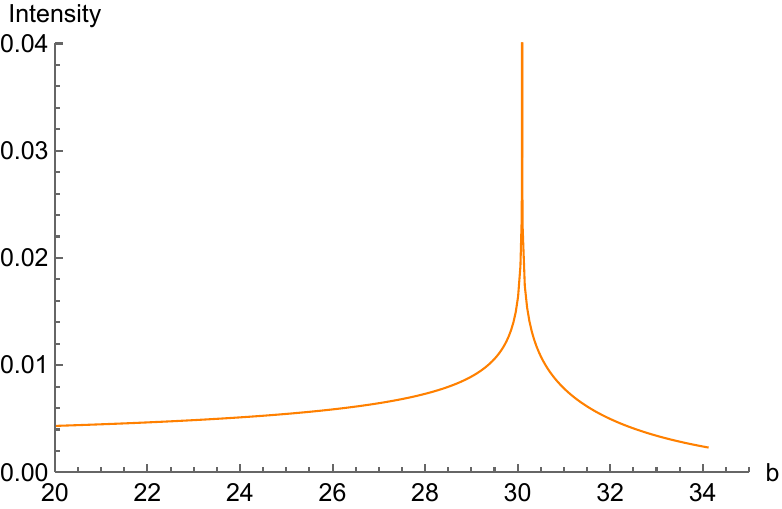}
        }
        \qquad
        \subfigure{
            \includegraphics[height=5.5cm]{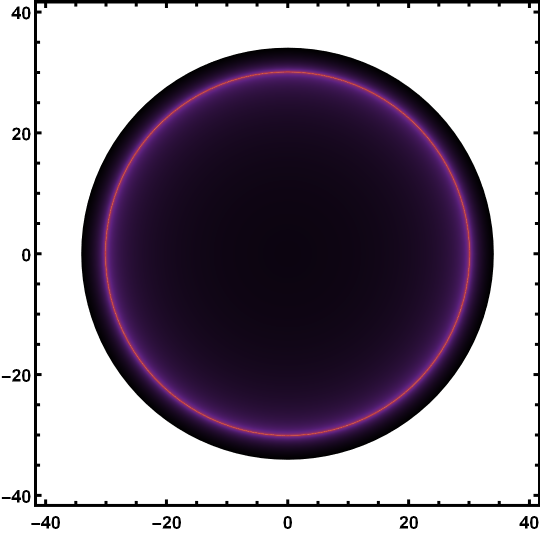}
        }
        \caption{The specific intensity and the black hole shadow with static spherical accretion for $a=0.6$, $\omega_q = -2/3$ and $\alpha = 0.01$.}
        \label{a=0.6}
    \end{center}
\end{figure}

As the parameter $a$ of string clouds converges to a maximum value, the radius of shadow approaches the cosmological horizon. When $\omega _q=-2/3$ and $\alpha =0.01$, we get $a<0.612158$. From the Table. \ref{Table 2}, we can see that $r_c=34.1421$ and $b_{ph}=30.0948$ when $a=0.6$. The observed specific intensity and shadow for $a=0.6$ is drawn in Fig. \ref{a=0.6}. Most of the specific intensity concentrates around $b=b_{ph}$. Moreover, from the profile of shadow in Fig. \ref{a=0.6}, it can be observed that all regions are relatively dark except for the vicinity of $b=b_{ph}$.

\begin{figure}[H]
    \begin{center}
        \subfigure[Intensity]{
            \includegraphics[width=6.5cm]{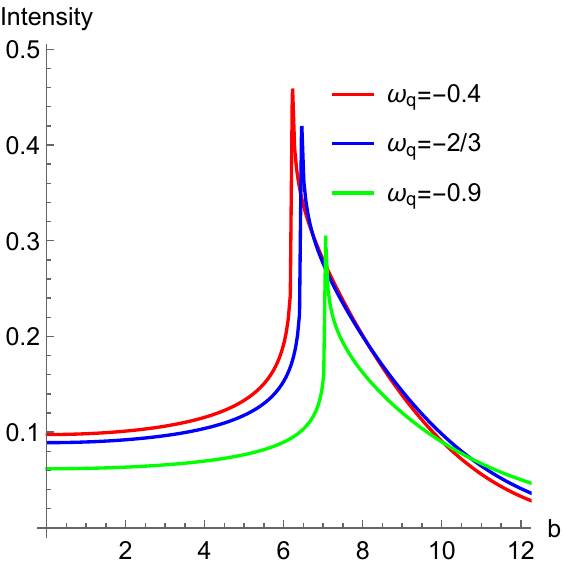}
        \label{inq}}
        \qquad
        \subfigure[$\omega _q=-0.4$]{
            \includegraphics[width=6.5cm]{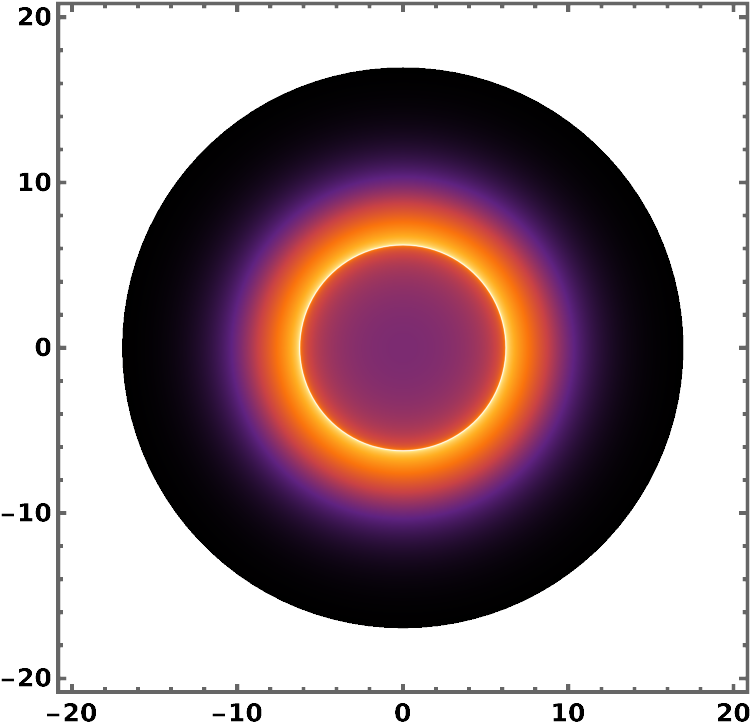}
        \label{Shadow-04}}\\
        \subfigure[$\omega _q=-\frac{2}{3}$]{
            \includegraphics[width=6.5cm]{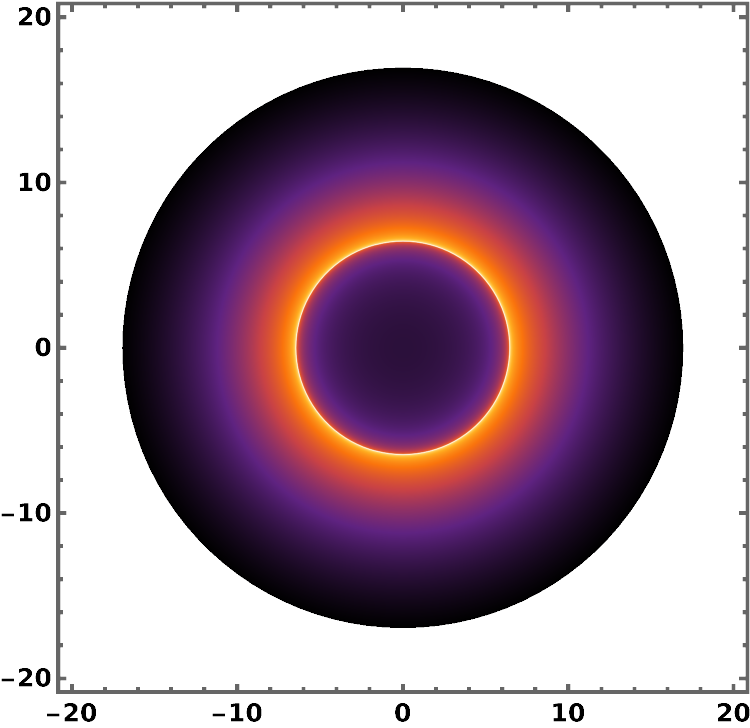}
        \label{Shadow-0667}}
        \qquad
        \subfigure[$\omega _q=-0.9$]{
            \includegraphics[width=6.5cm]{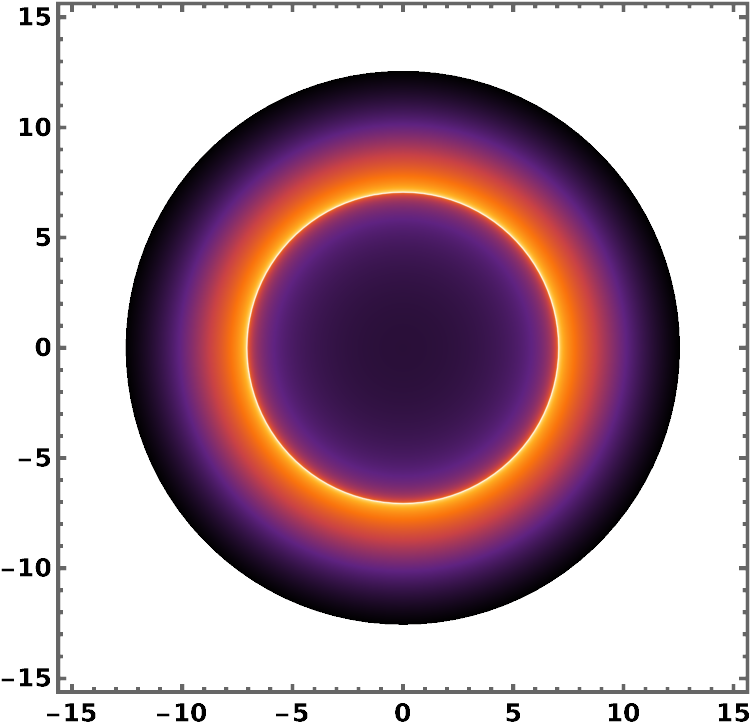}
        \label{Shadow-09}}
        \caption{The specific intensity for different $\omega_q$ in Fig. \ref{inq} and the black hole shadow with static spherical accretion for different $\omega_q$ in other images with $a=0.1$ and $\alpha = 0.01$.}
        \label{Shadow for Omega}
    \end{center}
\end{figure}

Using the same method, we investigated the specific intensities and the shadows for different $\omega_q$ with $a=0.1$ and $\alpha=0.01$ in Fig. \ref{Shadow for Omega}. From the image of specific intensity in Fig. \ref{inq}, it can be seen that $\omega_q$ would also affect the maximum value of specific intensity and the radius of the photon sphere. As $\omega_q$ decreases, the radius of the shadow increases. It can be seen from Table. \ref{Table} that $r_c=12.5872$ and $b_{ph}=7.06308$ for $\omega_q=-0.9$ which is small so we especially set the region in Fig. \ref{Shadow-09} to make $b\leqslant r_c$. The specific intensity inside the photon sphere does not vanish but has a small finite value, as some photons near the black hole have escaped \cite{Zeng:2020dco, Zeng:2020vsj}.

\section{Shadows and photon spheres with an infalling spherical accretion}

In this section, the accreting matter is free-falling in a static and spherically symmetric spacetime. When the angular velocity of the black hole is not considered, the accretion matter will only have radial velocity towards the black hole. We still employ Eq. (\ref{Intensity}) to investigate the shadow. The redshift factor in this model can be written as

\begin{align}
    g=\frac{p_{\alpha }u_o^{\alpha }}{p_{\beta }u_e^{\beta}},
    \label{g2}
\end{align}
where $p_\mu$ is the 4-momentum of photons, $u^{\mu}_o = (1, 0, 0, 0)$ is the 4-velocity of the distant observer, and $u^{\mu}_e$ is the 4-velocity of the photons emitted from the accretion. The matter which emits the photons at different locations has different radial velocities due to the tidal force from the black hole. In the infalling accretion, the $u^{\mu}_e$ can be expressed as \cite{Jaroszynski:1997bw}
\begin{align}
    u^{t}_e = \frac{1}{A(r)},\quad u^{r}_e = -\sqrt{\frac{1-A(r)}{A(r)B(r)}},\quad u^{\theta}_e = u^{\phi}_e =0.
    \label{ut}
\end{align}

From Eq. (\ref{Metric of spherical black hole}), we simply set $A(r) = f(r)$, $B(r) = 1/f(r)$ \cite{Zeng:2020dco}. As $p_t$ is a constant and $p_{\alpha}p^{\alpha}=0$, we obtain
\begin{align}
    p_r = \pm p_t\sqrt{\frac{1}{f(r)}(\frac{1}{f(r)} - \frac{b^2}{r^2})}.
    \label{pr}
\end{align}

Combining Eqs. (\ref{g2}), (\ref{ut}) and (\ref{pr}), the redshift factor can be expressed as
\begin{align}
    g=-\frac{f(r)}{\sqrt{1-f(r)} \sqrt{1-\frac{b^2 f(r)}{r^2}}-1}.
    \label{gb2}
\end{align}

\begin{figure}[H]
    \begin{center}
        \subfigure[Intensity for $a=0.1$]{
            \includegraphics[width=6.5cm]{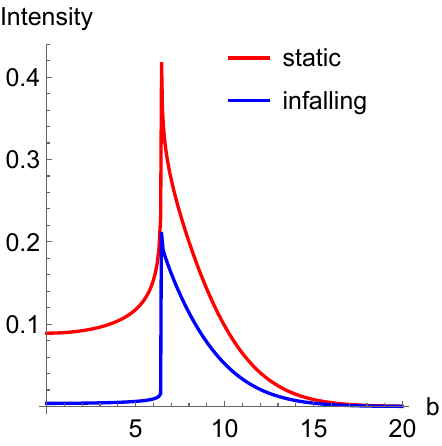}
        \label{ina01}}
        \qquad
        \subfigure[$a=0.1$]{
            \includegraphics[width=6.5cm]{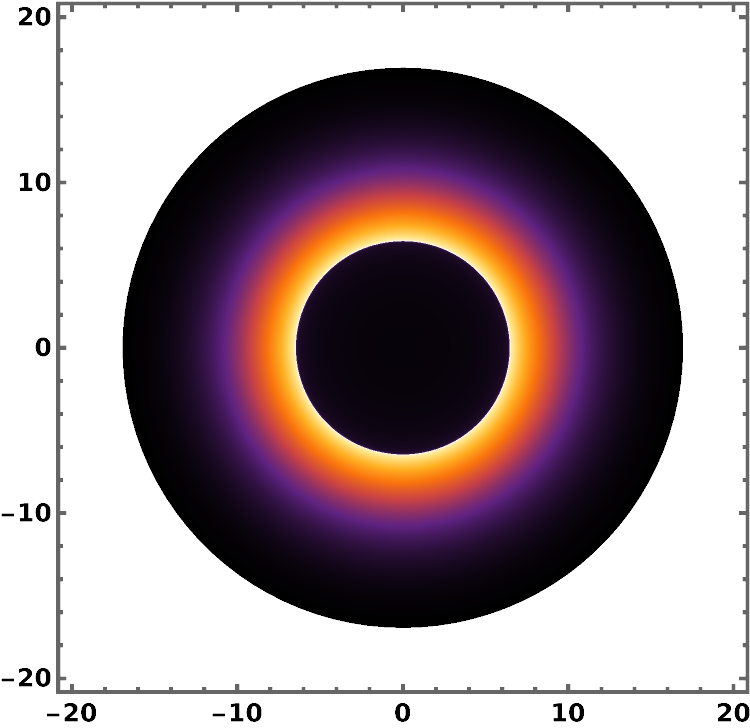}
        \label{insha01}}\\
        \subfigure[Intensity for $a=0.3$]{
            \includegraphics[width=6.5cm]{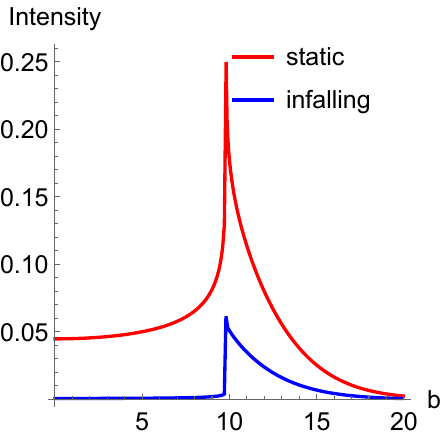}
        \label{ina03}}
        \qquad
        \subfigure[$a=0.3$]{
            \includegraphics[width=6.5cm]{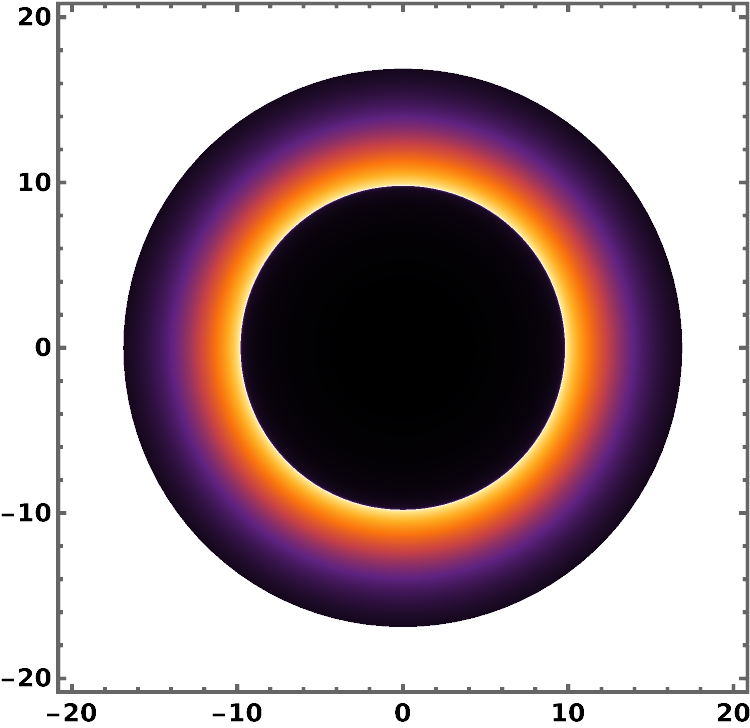}
        \label{insha03}}
        \caption{The specific intensity with static and infalling spherical accretion for $a=0.1$ and $a=0.3$ in Fig. \ref{ina01} and Fig. \ref{ina03} respectively. And the profile of black hole shadows with an infalling spherical accretion for $a=0.1$ and $a=0.3$ in Fig. \ref{insha01} and Fig. \ref{insha03} respectively with $\omega_q=-2/3$ and $\alpha=0.01$.}
        \label{Graphic of infalling for a}
    \end{center}
\end{figure}

Substituting Eqs. (\ref{jv}), (\ref{lpp}) and (\ref{gb2}) into Eq. (\ref{Intensity}) to get a function of specific intensity that are determined by the impact parameter $b$, then we can plot the black hole shadow with an infalling spherical accretion. The shadows for $a=0.1$ and $a=0.3$ are drawn in Fig. \ref{insha01} and Fig. \ref{insha03} respectively with $\omega_q=-2/3$ and $\alpha=0.01$. The interior of the shadows with an infalling spherical accretion will be darker than that with the static spherical accretion.  Due to the radial velocity of the accreting matter, the frequency of the photons we received reduces and the profile of shadow gets darker. The same conclusion can be shown from Eq. (\ref{gb2}).   

We are also interested in the difference for the specific intensity between static and infalling spherical accretion. With $\omega_q=-2/3$ and $\alpha=0.01$, we plot the graph of specific intensity with static and infalling spherical accretion for $a=0.1$ and $a=0.3$ in Fig. \ref{ina01} and Fig. \ref{ina03} respectively. As analyzed above, the blue line is much lower than the red line in $b<b_{ph}$. With $b$ increasing, the specific intensity $I(b)$ with either static or infalling spherical accretion gradually converges, because the emitted photon with larger impact parameter $b$ will move to a smaller average radial position in the integral along the geodesic, and the effect of the redshift factor diminishes. When $a=0.3$ in Fig. \ref{ina03}, the difference between the blue line and the red line also increases since the observers and black holes get closer, which in turn leads to a larger redshift factor $g$ from Eq. (\ref{gb2}).

\begin{figure}[H]
    \begin{center}
        \subfigure[Intensity for $\omega _q=-0.5$]{
            \includegraphics[width=6.5cm]{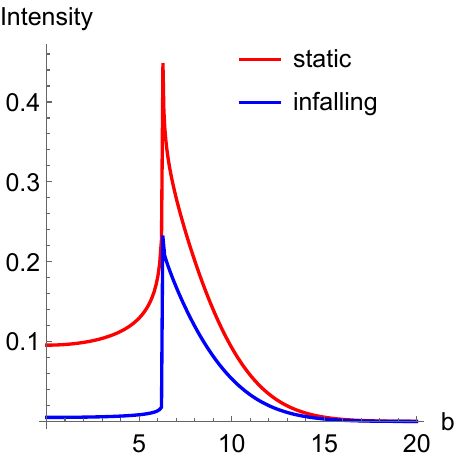}
        \label{ino-05}}
        \qquad
        \subfigure[$\omega _q=-0.5$]{
            \includegraphics[width=6.5cm]{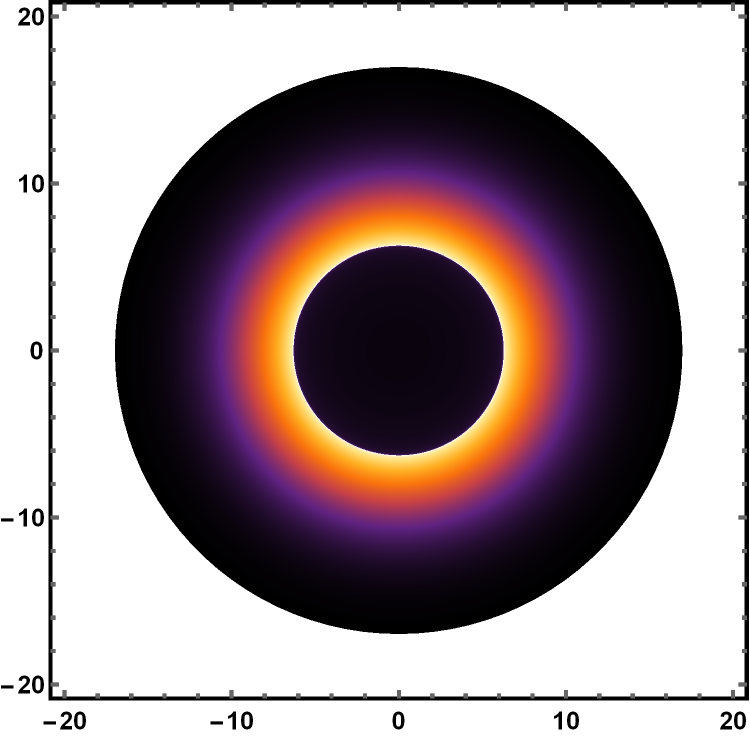}
        \label{insho-05}}\\
        \subfigure[Intensity for $\omega _q=-0.9$]{
            \includegraphics[width=6.5cm]{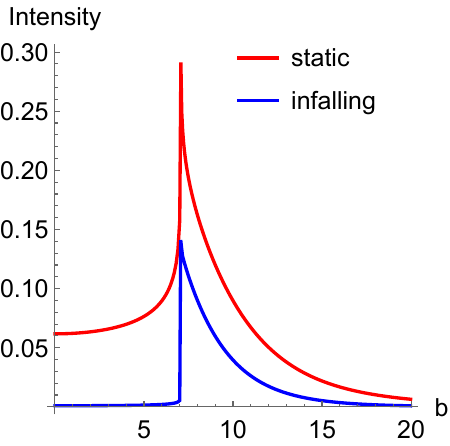}
        \label{ino-09}}
        \qquad
        \subfigure[$\omega _q=-0.9$]{
            \includegraphics[width=6.5cm]{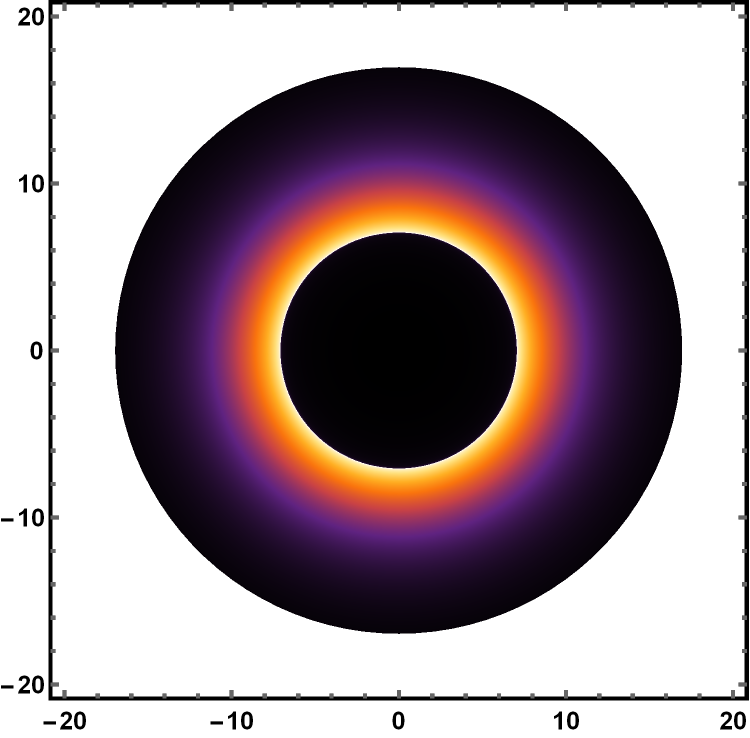}
        \label{insho-09}}
        \caption{The specific intensity with static and infalling spherical accretion in Fig. \ref{ino-05} and Fig. \ref{ino-09} respectively. And the profile of black hole shadows with an infalling spherical accretion in Fig. \ref{insho-05} and Fig. \ref{insho-09} for $\omega_q=-0.5$ and $\omega_q=-0.9$ respectively with $a=0.1$ and $\alpha=0.01$.}
        \label{Graphic of infalling for omega}
    \end{center}
\end{figure}

In the same way, we further investigate how different $\omega_q$ impact the specific intensity with an infalling spherical accretion in Fig. \ref{Graphic of infalling for omega}. From Table. \ref{Table}, the radius of shadow $b_{ph}$ varies very slowly with $\omega_q$, which causes the bright rings in Fig. \ref{insho-05} and Fig. \ref{insho-09} to have a similar size. For both static and infalling spherical accretion, the specific intensity varies for different $\omega_q$ because the length of the optics path varies with $\omega_q$.

Either with static or infalling spherical accretion, the radius of the shadow is the same in both Fig. \ref{Graphic of infalling for a} and Fig. \ref{Graphic of infalling for omega}, in other words, the peak of the images of specific intensity are all located at $b=b_{ph}$. We can conclude that the model of the accretion we considered in this paper only affects the value of the specific intensity but not the profile. Because the radii of the shadow and photon sphere is actually the intrinsic properties of spacetime. 

\section{conclusion and discussion}
We firstly derived the black hole solution combining quintessence and string clouds, and used the Euler-Lagrangian equation to obtain the geodesics of photons. By analyzing the influence of parameters on photons trajectories, we learn that the radius of the photon sphere $r_{ph}$ is non-monotonic with $\omega_q$ decreasing, which is consistent with the previous results \cite{Zeng:2020vsj}. When $a$ increases, the radius of the event horizon $r_h$, cosmological horizon $r_c$, photon sphere $r_{ph}$, the corresponding impact parameter $b_{ph}$ all vary monotonically. Then, we discussed the range of $a$ in different $\omega_q$. As the photons are deflected, it forms the shadow of a black hole.

Moreover, we studied the black hole shadow with static and infalling spherical accretions respectively. The backward ray shooting method helps us obtain the specific intensity $I(b)$ from a static observer. For the spherical accretion, we assume that the density of light sources follows a normal distribution affected by the coefficient $\gamma$. And the photons emitted are not strictly monochromatic, whose frequencies conform to the Gaussian distribution with a central frequency $\nu_*$ and width $\Delta\nu$. The photon emissivity is written as $j(\nu_e)$. Using the backward ray shooting method and the model of the photon emissivity, we plotted the shadow and photon sphere with static and infalling spherical accretions. Due to the Doppler effect, the interior of the shadow with an infalling spherical accretion will be darker. Different $\omega_q$ and $a$ would change the size and intensity of the shadow. We investigated the emissivity with different coefficients $\gamma$ and found that the intensity will decrease as $\gamma$ increases.

In the presence of the cosmological horizon, the range of $a$ is not arbitrary, which needs to satisfy $r_c>b_{ph}$. As if $r_c<b_{ph}$, then all photons emitted from the observer will not be able to escape from the black hole, which causes the specific intensity to be little.

Actually, there are some flaws in the backward ray shooting method, which means when we compare the specific intensity observed at different distances from the black hole, the intensity observed by the distant observer is stronger than that of the closer observer in Fig. \ref{ina} and Fig. \ref{inq}. When the cosmological horizon is not considered, in other words, when the observer's position is fixed, the ray shooting method may appear more realistic. We can also make it more practicable by assuming the observer's position changes with the size of cosmological horizon $r_c$, such as improving the accretion model or applying a more realistic frequency distribution function.

The EHT Collaboration portrays M87* and claims that the observation supports General Relativity. We expect this to bring insights and implications for the quintessence dark energy model and the string theory from the observations of the black hole shadow  in future astronomical observation projects.

\begin{acknowledgments}
We are grateful to Peng Wang, Hanwen Feng, Yuchen Huang, Qingyu Gan and Wei Hong for useful discussions. This work is supported by NSFC (Grant No.11947408 and 12047573).
\end{acknowledgments}

\bibliographystyle{unsrt}
\bibliography{egbib}

\end{document}